# Photoemission electron microscopy of localized surface plasmons in silver nanostructures at telecommunication wavelengths

Erik Mårsell[1], Esben W. Larsen [1], Cord L. Arnold[1], Hongxing Xu[1], Johan Mauritsson[1], and Anders Mikkelsen[1,a]

[1]Department of Physics, Lund University, P.O. Box 118, 22 100 Lund, Sweden

(Dated: 29 January 2015)

We image the field enhancement at Ag nanostructures using femtosecond laser pulses with a center wavelength of 1.55 micrometer. Imaging is based on non-linear photoemission observed in a photoemission electron microscope (PEEM). The images are directly compared to ultra violet PEEM and scanning electron microscopy (SEM) imaging of the same structures. Further, we have carried out atomic scale scanning tunneling microscopy (STM) on the same type of Ag nanostructures and on the Au substrate. Measuring the photoelectron spectrum from individual Ag particles shows a larger contribution from higher order photoemission process above the work function threshold than would be predicted by a fully perturbative model, consistent with recent results using shorter wavelengths. Investigating a wide selection of both Ag nanoparticles and nanowires, field enhancement is observed from 30% of the Ag nanoparticles and from none of the nanowires. No laser-induced damage is observed of the nanostructures neither during the PEEM experiments nor in subsequent SEM analysis. By direct comparison of SEM and

[a] Electronic mail: anders.mikkelsen@sljus.lu.se

PEEM images of the same nanostructures, we can conclude that the field enhancement is independent of the average nanostructure size and shape. Instead, we propose that the variations in observed field enhancement could originate from the wedge interface between the substrate and particles electrically connected to the substrate.

**I. INTRODUCTION**

The field of nano-optics has experienced a tremendous development in recent years due to both the availability of new tailored nanostructures and the development of highly advanced imaging and spectroscopy tools for studying these effects [1-6]. One imaging concept that has seen significant results lately is the coupling of state-of-the art laser technology with photoemission electron microscopy (PEEM). This approach combines the ability to tailor the temporal and spectral structure of laser fields with the high spatial resolution of electron microscopy [7-12]. In addition, the strong dependence on the near-field strength and the electron emission pathways provides a high sensitivity to a variety of surface properties [13,14]. Finally, the near-field character of the technique allows for the study of dark modes and other effects that cannot be observed using conventional far-field spectroscopic methods [2,15,16].

PEEM for studying field enhancement effects relies on the non-linear emission of electrons by photons with energies below the workfunction threshold (3-5 eV). Typically this has been performed using lasers with wavelengths of 400 nm or 800 nm, which already at moderate field intensities will induce two or three photon photoemission. This can be done in the perturbative regime, where the total photoelectron yield scales with the field intensity to the power of two or three depending on the laser wavelength and the surface work function, and the photoelectron spectrum drops off exponentially with increasing energy. Extension of laser-based PEEM imaging to lower photon energies allows near-field imaging of light–nanostructure interactions in the technologically important telecommunication wavelength region. Field enhancement effects at these wavelengths can be used to boost optical communication or enhance conversion to electrical signals [17,18]. Indeed, wedges and edges as a sources of plasmon concentration for propagation at these wavelengths have been studied thoroughly in recent years [15,19]. In theoretical models, strong subwavelength field enhancements are observed indicating the relevance of imaging on a spatial scale far below the free-space wavelength of the light. Many Ag and Au nanostructures, which are the common model structures for plasmonic studies, have localized plasmonic resonances at these wavelengths with field concentration at wedges or crescents [20-23].

Using 0.8 eV radiation for photoemission experiments presents several new opportunities and challenges. Reported work functions for Ag are in the 4.1-4.7 eV region, meaning that the photon energy is below one fifth of the work function of the material. This further implies that the intensities required for imaging are likely to be beyond the perturbative regime of direct n-photon photoemission via virtual states. Other mechanisms such as field emission [13], thermal emission [24], and defect-mediated emission [25], as well as combinations of the above [26], have been proposed to generate electron emission by optical pulses. Regardless of the exact mechanism involved, the laser-induced emission of an electron from Ag will due to energy conservation require at least six 0.8 eV photons, and the photoelectron yield is expected to depend non-linearly on the local near-field intensity. Going to longer wavelengths and thus lower photon energies generally reduce the electron emission probability and at some point the loss of photon energy to phonons compared to electron emission will result in a situation where the nanostructures will start to melt or ablate before any electron emission is observed. This is not an unreasonable concern for the silver structures investigated in the current study. This type of nanostructures has been used in a large number of plasmonic studies [27,28] and are highly relevant also for imaging of attosecond phenomena using PEEM [29,30]. However, they are known to in some instance be melting even at very moderate temperatures or in the presence of plasmonic hot spots [31].

In this work, we have performed PEEM using femtosecond laser pulses at the technologically important wavelength of 1550 nm. The samples consist of a variety of rationally synthesized Ag nanostructures deposited on a Au film, giving rise to localized surface plasmon resonances across a wide spectrum. The nonlinear photoemission process calls for a high-intensity light field. By using an optical parametric amplifier together with a high power laser system, pulse energies of around 0.7 mJ are possible with pulse lengths of 30 fs. In the current experiments, the laser power was tuned down in order to avoid space charge effects. One important question is whether the high intensities needed for imaging would exceed the damage threshold of the delicate nanostructures. Further, the extreme sensitivity of the non-linear photoemission process might result in so strong fluctuations in the number of emitted electrons that imaging becomes practically impossible. Our experiments show that PEEM imaging is still possible, despite these concerns.

**II. EXPERIMENT**

The laser system used in the experiments is based on a Ti:Sapphire regenerative amplifier system delivering 20 fs pulses with up to 5 mJ energy and 800 nm center wavelength at a repetition rate of 1 kHz. The pulses are sent into a high-energy TOPAS-Prime (Travelling-wave Optical Parametric Amplifier of Superfluorescence) from Light Conversion Ltd for tunable frequency conversion over a wide range of output frequencies. The TOPAS consists of a sapphire plate used for generation of white light seed, followed by three amplification stages where a selected part of the white light spectrum is amplified through optical parametric amplification in BBO crystals. The output is two linearly polarized IR pulses with wavelengths that can be tuned from 1160 to 1600 nm and from 1600 to 2600 nm respectively, a duration around 30 fs, and a total converted energy up to 1.7 mJ per pulse. A typical spectrum of the TOPAS output is shown in Fig 1b. A 1 m focal length lens loosely focuses the beam onto the sample at an angle of 65 degrees with respect to the normal. In the experiments reported in this paper, we estimate the peak intensity incident on the sample to be on the order of $5*10^9$ W/cm$^2$. Compared to previous PEEM studies using 800 nm light, this intensity is in the high part of the reported ranges [11-13], which is expected due to the lower photon energy. The laser beam is s-polarized, i.e. the electric field vector lies in the sample plane.

The PEEM is a commercial instrument from Focus GmbH, located in an ultra-high vacuum chamber. It accelerates the photoelectrons using a 10-15 kV voltage and forms an image of the photoelectrons using an electrostatic lens system. The PEEM is equipped with a high-pass imaging energy filter (IEF) for spectroscopic analysis. The experimental setup is schematized in Fig. 1a. For all images used in the analysis below, the laser intensity is tuned to a level where no significant space charge effects are observed. The instrument is also equipped with a Hg discharge lamp for UV-PEEM using continuous-wave illumination at 4.9 eV.

The sample is made from colloidal Ag nanowires and nanoparticles made using a polyol process, and dispersed in ethanol solution [32,33]. In the polyol process, poly-(vinyl pyrrolidone) (PVP) preferably attaches to the Ag(001) facets of nanocrystals, thus favors one-dimensional growth. A droplet of the solution is placed on a 50 nm thick Au film on Si and blow-dried after 30 s. The resulting sample has a mixture of Ag nanowires with diameters of around 150 nm and lengths of a few tens of microns, and Ag nanoparticles with a variety of shapes, with average sizes of 100-150 nm.

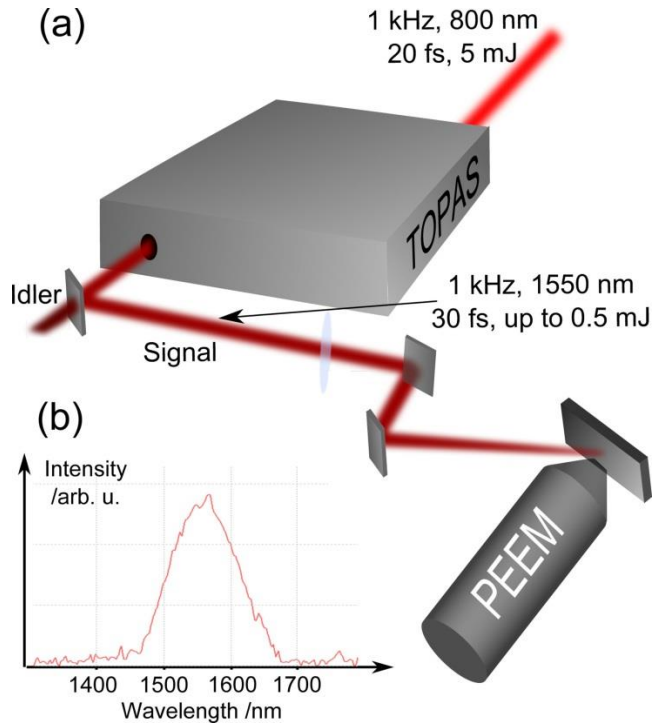

Figure 1. (a) Schematic of the experimental setup. At the output of the TOPAS, the signal and idler beams propagate collinearly. The signal is reflected off a dichroic beamsplitter, which transmits the idler. The s-polarized laser beam is then loosely focused onto the sample. (b) Recorded spectrum of the output of the TOPAS when operated at 1550 nm. The spectrum shows

that the combination of optical parametric amplification and spectral filtering by the dichroic beamsplitter gives a nicely bell-shaped spectrum centered around 1550 nm.

## III. RESULTS AND DISCUSSION

An SEM image of part of a typical sample is shown in Fig. 2a, showing a collection of nanowires and nanoparticles. The nanoparticles have a variety of sizes and shapes, and have distinct crystalline facets. To further characterize the surfaces of the Ag nanoparticles, we perform scanning tunneling microscopy (STM) measurements on a similar sample as used for the PEEM measurements. First, the Au film is imaged (Fig. 2b). The STM investigations show that the size of the grains in the polycrystalline Au film is on the order of 50 nm, and that the RMS roughness of the surface is 2 nm (see inset histogram of Fig. 2b). A collection of nanoparticles can be seen in an overview image (Fig. 2c). In this way, we can locate and identify the different nanostructures on the sample before zooming in and performing more detailed studies. We can proceed to both find and image the Ag nanostructures and their top surfaces. STM images of the nanowires show a broad, smooth cylindrical shape (Fig. 2d). While the top part of the nanowire shape can be trusted one should be aware that it represents a convolution with the tip resulting in a larger width of the wire. The high-resolution STM images of Fig 2e show that we can get atomic resolution on top of the Ag particles without any in-vacuum treatment such as heating or exposure to reactive gases to remove oxides or organic residues. A height profile is measured along the red line in Fig. 2e, as seen in Fig. 2f-g. The smooth oscillations correspond to single atomic distances with a spacing of about 2.7 Å which is consistent with an unreconstructed low index Ag metal surface. Thus our measurements show that the particles and wires are generally very smooth with sub-nanometer roughness and that at least some of the particles have not sulfurized or oxidized. Our measurements cannot exclude the existence of a homogeneous monolayer of PVP from the synthesis process on the nanowires [34-36]. On the nanoparticles though, the STM investigations indicate a clean, metallic surface. From the synthesis mechanism, it is known that the (111) surfaces of the nanoparticle have a much lower affinity for PVP than the (001) surfaces of the nanowires, which is in good agreement with our STM measurements. Thus we can conclude that some of the Ag particles with low index facets will likely be in direct contact with the Au substrate, while the nanowires could have a PVP layer in between them and the Au surface.

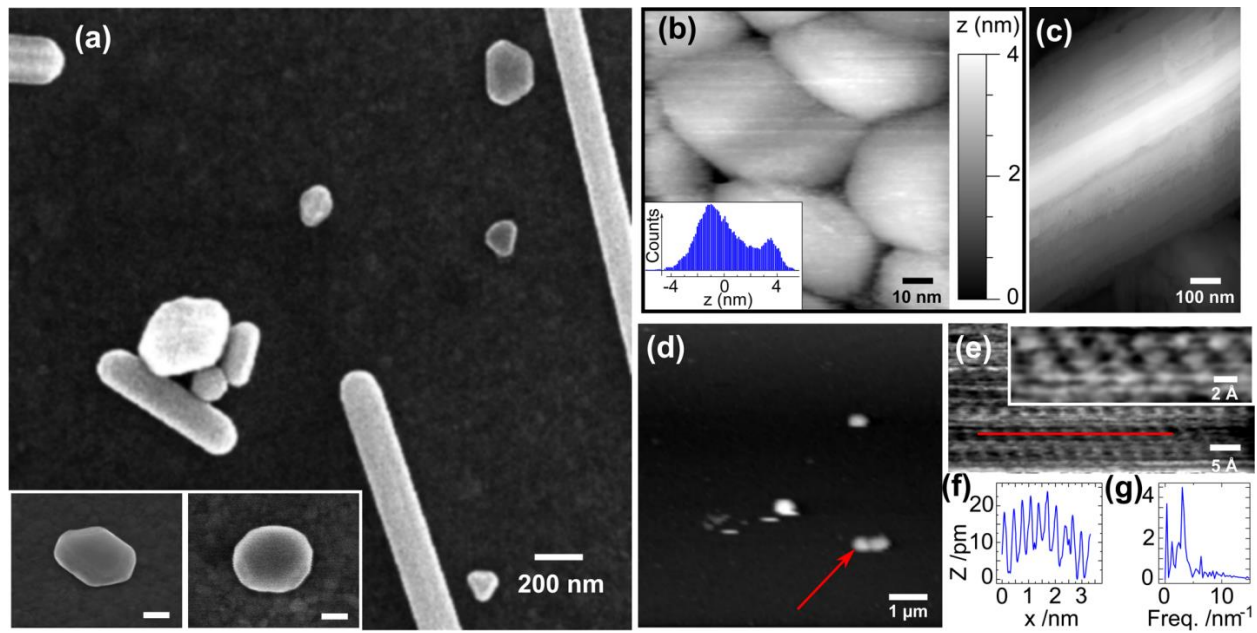

Figure 2. a) SEM image of a collection of nanoparticles and nanowires similar to what would typically be found on samples used for the PEEM experiments. Note the clear faceting and the varying size and shape of the Ag particles. The two insets show zoomed in SEM images of two particles with different shapes, acquired at a 30° tilt angle. Inset scalebars are 100 nm. b) 100 x 100 nm$^2$ STM image (+0.1 V, 700 pA) of the polycrystalline Au film. The crystal grains in the film can be clearly seen. The z scale (right) shows corrugations on the order of a few nm. Inset:

Histogram sampled from a 500 x 500 nm$^2$ area of the substrate, showing the height distribution. The RMS roughness of the area is 2.1 nm, and the difference between the highest and the lowest pixel is 11 nm. c) 6 x 8 µm$^2$ overview image (-2.0 V, 10 pA) of a collection of Ag particles. d) 700 x 1000 nm$^2$ STM image (-1.0 V, 100 pA) of a nanowire. e) 5 x 2.5 nm$^2$ image (-0.2 V, 700 pA) of a plateau on top of the particle marked with an arrow in d). Some atomic rows can be clearly resolved. The inset (+0.1 V, 900 pA) shows a small area of the same particle where single atoms could be imaged. This resolution is reached even though no heat treatment or other in-vacuum cleaning of the sample has been performed. f) Height profile measured along the red line in e. g) Fourier transform of the height profile, showing a distinct peak corresponding to a period of 3.3 Å. The line cut is taken at an angle of 35° with respect to the direction perpendicular to the atomic rows, corresponding to a distance between the rows of 2.7 Å.

Typical PEEM results for this system are displayed in Fig. 3. In Fig 3a we show an overlay image of the signal recorded with the 1550 nm laser as excitation source (in red), with a UV-PEEM image of the Au film with Ag nanowires and nanoparticles (blue). This allows us to identify each nanostructure in the image, to later correlate with scanning electron microscopy (SEM) images of the very same areas, which is seen in Fig. 3b. It can immediately be observed that no photoemission from the Ag nanowires can be detected under these conditions, as opposed to many of the Ag nanoparticles. Two further observations can also be made: First, all bright spots in the PEEM image can be correlated with a Ag nanoparticle or an assembly of particles. We also observe a number of Ag particles that do not emit electrons at a detectable level. Looking at the overview images such as Fig. 3 we can conclude that roughly 30% of the nanoparticles appear in the PEEM images, meaning in practice that they enhance the field to a similar level. Due to the non-linear response, very small changes in the field can be observed. For example, in the perturbative regime the 6-photon photoemission yield would scale as the $6^{th}$ power of the near-field intensity – thus a 1% increase in the field amplitude will correspond to a 13% increase in the photoemission yield. Even if the photoemission process in our experiments is not fully described by perturbation theory, as will be discussed later, we can conclude that at least 6 photons (4.8 eV) will be needed for each photoelectron, as reported workfunctions of Ag are in the range of 4.1-4.7 eV and the workfunction of Au is around 5.1 eV. We can thus expect the photoemission yield to depend very sensitively on the local field enhancement.

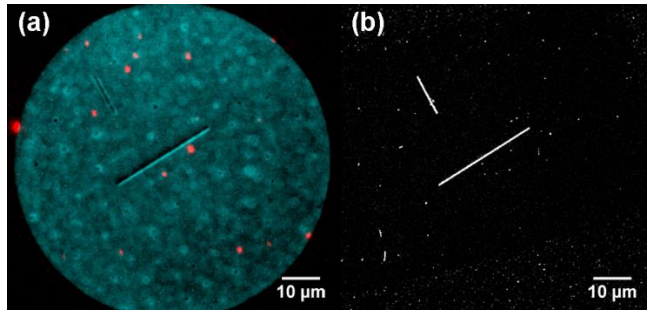

Figure 3. a) UV-PEEM image of Au film with Ag wires and particles. The appearance of an assymetric dark and bright part of each particle and wire is due to the light coming in at an angle of 25 degress with respect to the substrate (from the lower right as seen in the image), which gives rise to a shadowing effect. The 1550 nm PEEM image (acquisition time 8 s) of the same area is colored red and overlaid on top of the UV-PEEM image. The electron emission is confined to a few hot spots. The laser light is incident from the left in the image. b) SEM image of the same area.

Comparing the PEEM images recorded by the UV light, 1550 nm PEEM images and SEM images from the same areas we are able to further investigate which particles that give a detectable electron emission when illuminated by the laser pulses. The SEM images show that there are no clear differences between particles that respond to the light and those that do not. In both cases, particles range in diameter from 80 nm to 200 nm. For some of the larger particles, an asymmetry can be seen in the PEEM images indicating that the photoemission is not homogeneously distributed across the particle. No degradation of the particles is seen to occur in the PEEM during the experiments, which is also confirmed in the subsequent SEM analysis. The localized laser-induced electron emission from only a fraction of the particles and no indication of heating or ablation effects agree well with local plasmonic field enhancements expected to be found in the Ag–Au materials system.

To further study the response we performed energy filtered PEEM imaging. In this way, we can compare the shape of the electron spectrum with the total number of emitted electrons for each nanoparticle. The spectra in general show two regions that can be approximated with straight lines in a semi-logarithmic plot: a plateau region with a small slope, followed by a cut-off region with a steeper slope. The width of this plateau region varies among the particles, but is on the order of 1–3 eV, corresponding to above-threshold ionization [37] by at least three photons. Fig. 4 shows a region of the sample imaged with Hg lamp (a) and with 1550 nm laser pulses (b). The photoelectron spectra are plotted on a semi-logarithmic scale in Fig. 4 (c) for 4 different spots from (b), and in each spectrum the two regions can be identified. We note that the two most intense spots, marked 2 and 3 in the figure, are also the ones with the widest plateau in the photoelectron spectrum, roughly 2-3 photon energies wide, as expected for non-linear photoemission. The shape of the spectra corresponds well with studies of laser-assisted photoemission at moderate intensities such as the one by Aeschlimann et al. [38] and work by Schertz et al [13]. The flat part of the spectrum indicates that the photoemission process is not fully perturbative, since this would have yielded a spectrum that was rapidly decreasing even at low energies, and that would show kinks separated by the photon energy (see e.g. [39]). Another possible source of spectrum broadening is space charge effects [40]. However, space charge effects would also be observed as a blurring of the images, and these experiments were performed at intensities where no such blurring of the hot spots could be observed. Furthermore, by counting single electron events on the double MCP of the PEEM and comparing the number of counts on the CCD, we estimate that the brightest hot spots of the PEEM images in this paper consist of approximately 1000 electrons. With an acquisition time of 60 s and a repetition rate of 1 kHz, this corresponds to much less than one electron per laser pulse. Heating of the electron gas by the strong electromagnetic field has also been considered for electron emission. This mechanism is favoured for long pulses and small particles, since a higher density of defects gives a larger probability for electron scattering [24,25]. However, we cannot completely exclude the possibility of electrons undergoing electron-electron scattering within the duration of the laser pulse [41]. The shape of the electron spectrum is in good agreement with the observations of Schertz et al [13] observed using 800 nm laser light. In this case the emission was explained via field emission and modelled via a dynamic Fowler–Nordheim equation, followed by ponderomotive acceleration in the local plasmonic field. Generally, one can envision both pure perturbative multi-photon photoemission, tunneling field emission or some combination of the two [26], which can be hard to exactly quantify. Regardless of the exact mechanism, we can expect a clearly non-linear photoemission process simply because energy conservation requires at least 6 photons to be absorbed per emitted electron. This is confirmed by fluctuating intensities both from nanoparticle to nanoparticle and over time during the measurements.

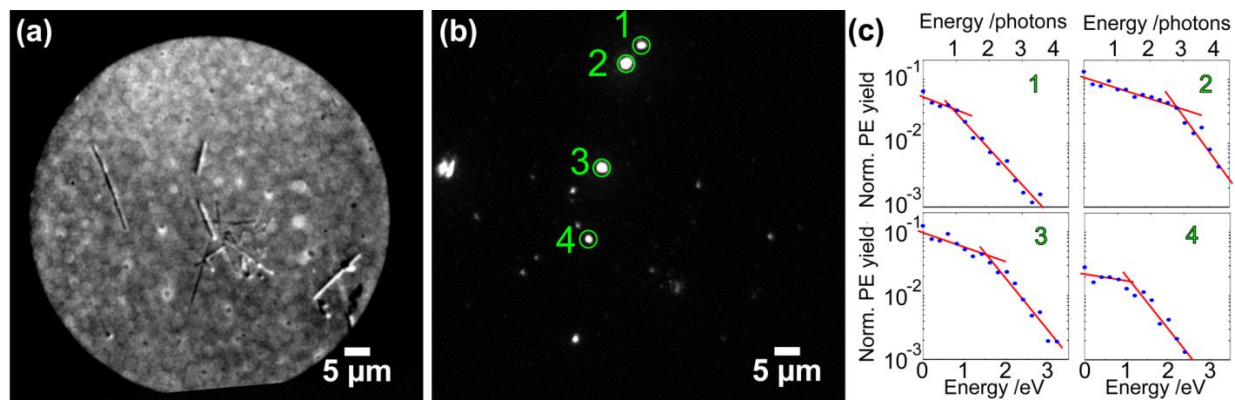

Figure 4. (a) Hg lamp image of a collection of nanowires and nanoparticles. (b) 1550 nm PEEM image (acquisition time 60 s) of the same area. (c) Photoelectron spectra (acquisition time 60 s per image) from the 4 spots labeled in (b). The lower x-axis shows the electron kinetic energy in eV, while the upper x-axis shows it in number of 0.8 eV photons. The y-axis is logarithmic and shows the normalized photoelectron yield. The red lines are added as a guide to the eye to mark the two regions in the spectra.

Reproducibility and noise are relevant concerns for the detection of photoelectrons from non-linear photoemission, where small variations of the laser field can strongly influence the intensities observed in the PEEM images. A small area was analyzed in a similar way as the overview image in Fig. 4, and is displayed in Fig. 5. The inset (Fig. 5c) shows an SEM image of the one particle in the area that gives a detectable photoemission. The photoelectron spectrum from the particle is shown in Fig. 5d. The two data sets represent measurements of the same single nanoparticle repeated twice. As shown also in Fig. 4, the spectrum consists of two parts that each can be approximated with a straight line. The main difference between the two measurements is the intersection point between these lines: the first measurements show a plateau region in the spectrum that reaches approximately 0.4 eV higher than for the second measurements. This can be explained by laser drift giving rise to a lower intensity of the IR radiation, as is also indicated by the total photoelectron yield from the spot in the first measurement being ~20% higher than in the second measurement. The higher photoemission intensity corresponds to a higher surface electric field, which can also lead to more transferred energy to the emitted electrons and therefore push the plateau region of the spectrum towards higher energies. This is again consistent with electron emission beyond the pertubative regime of direct multiphoton photoemission [13,38,41].

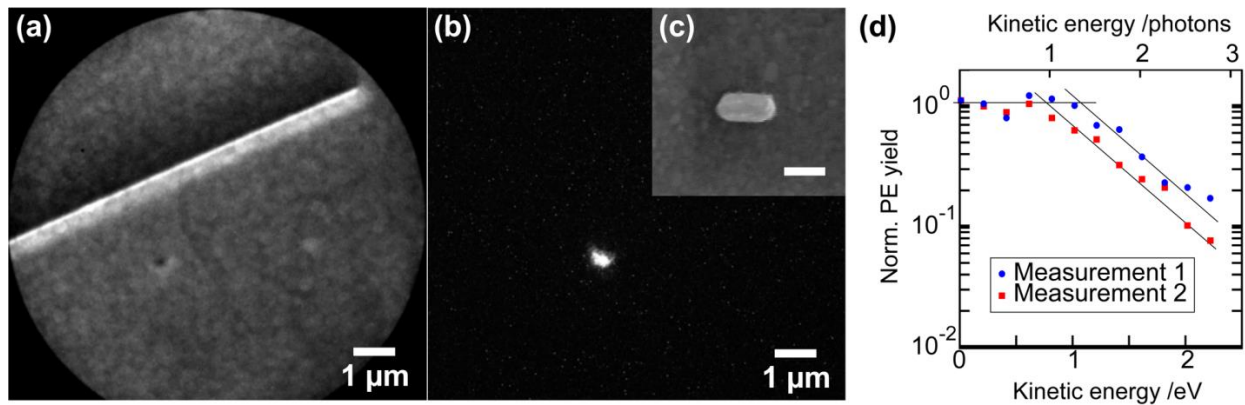

Figure 5. a) Close-up UV-PEEM image of a part of the area shown in Fig. 3. b) 1550 nm PEEM image (acquisition time 24 s) of the same area, showing emission from the particle located approximately 2 μm from the nanowire. Inset (c) shows an SEM image of the particle in question. Scalebar is 200 nm. d) Energy spectrum of the electrons emitted by the single particle (acquisition time 60 s per image). Data is shown for 2 different measurements, giving an estimate

of the stability of the setup. The lines are guides to the eye showing the two distinct regions in the spectrum. The first measurement shows a plateau that stretches approximately half a photon energy further than in the second measurement.

Ag nanoparticles with these sizes exhibit dipole resonances that mostly span the visible spectrum, suggesting that the 1.55 µm radiation is off-resonance. However, we anyway observe a field enhancement from some of these particles but not from others, strongly indicating that some type of resonance condition is met for specific particles. This enhancement is completely uncorrelated with the average size and shape of the nanoparticles (from direct comparison of SEM and laser PEEM images of the same 47 particles), which is similar to what has been observed previously using 2-photon PEEM studies of a similar system [42]. We suggest that the observed field enhancement occurs in wedge-like features at the substrate–particle interface, especially when these are conductively connected. This is in contrast to other situations [2,13,31] where there is a thin insulating barrier between particle and metallic substrate. The existence of an enhanced field in the region of the substrate-particle interface is verified by electromagnetic simulations using two different methods. The RF module of COMSOL Multiphysics is used for finite element method calculations of the electric field around a Ag particle with dimensions corresponding to the one in Fig. 5, on top of a Au substrate. Direct contact between the particle and the substrate is ensured by cropping the bottom 1 nm of the structure, resulting in a flat, finite contact area. Just like in the experiment, the excitation has an incidence angle of 65 degrees, a wavelength of 1550 nm, and a polarization in the plane of the substrate. The situation is shown schematically in Fig. 6a. Fig 6b-c show the resulting field enhancement in different projections of the 3D system. The electric field enhancement in the plane of the substrate is seen in Fig. 6c, which shows a field enhancement factor of approximately 5. These results are also confirmed by finite-difference time-domain calculations of a similar system. A movie showing the electric field as a function of time during and after excitation by a 30 fs pulse in a plane at the Au surface (i.e. the same projection as in Fig. 6c) is presented as Supplementary information [43].

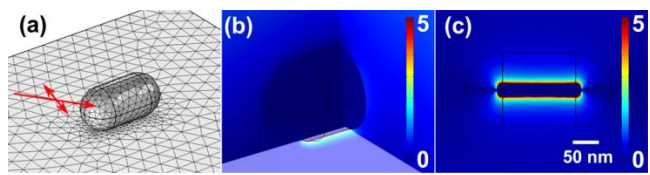

Fig. 6. Results of the FEM modeling. a) Sketch of the experimental situation and the mesh used for the calculation. b) Norm of the electric field in three different planes. c) Norm of the electric field at the Au surface.

With the non-linear detection scheme of our experiment, a field enhancement factor of 5 can dramatically increase the probability of photoemission. Small changes in the exact geometry of the substrate-particle interface can then explain the appearance of only some nanoparticles in the PEEM images. A competing explanation would be field concentration at edges and small irregularities of the nanoparticle surface. While this could give rise to differences in field strength at the surface, it does not explain why no electron emission can be detected from the nanowire ends. As we discussed above, our studies together with previous studies of polyol synthesized Ag nanostructures indicate that we can have a combination of structures with insulating layers (especially the nanowires which are stabilized with PVP on their surfaces), while the pure Ag surfaces found on the nanoparticles could well form a conductive path with the substrate. In similar systems with metallic particles in contact or separated by a small gap, the charge transfer plasmons occurring across the conductive junction have been shown to shift to wavelengths in the near-infrared in a region including 1550 nm, and significantly longer than the visible–regime plasmons observed when an insulating layer exists between the particle and a metallic substrate [44-46]. Returning at this point to the electron emission scheme proposed by Shertz et al, this also depends on acceleration of the electrons in the strong fields in a gap between the particle and the surface. In our case, without a complete gap between the particle and the substrate, we note that nonetheless there will be enhanced field lines in the wedge-like region of the particle–substrate interface, where the electrons are emitted. In contrast to the situation with an insulating gap, the wedge region at the conducting nanoparticle-substrate junction can have enhanced fields also for excitation by s-polarized radiation. The large variations in the detected signal can be explained by small differences in this wedge region, combined with the nonlinearity of the detection scheme which makes the PEEM signal extremely sensitive to the local field. We finally also note that our method is only dependent on the electromagnetic fields at the surface, and thus image both dark and bright plasmon modes, which is highly relevant for evaluating the plasmonic response of nanostructures.

**IV. CONCLUSION**

We have used PEEM to study near-field enhancement by Ag nanostructures using 1550 nm 30 fs laser pulses. The photon energy is less than one fifth of the work function of the material, and we observe electron emission up to approximately 3 photon energies above the vacuum level. The method is extremely sensitive to the local surface field enhancement, especially at the substrate-particle interface. This makes PEEM able to detect small differences in the field enhancement with high spatial resolution, differences that would be hard to observe with other existing methods. At intensities just below the space charge threshold, we note that the energy spectrum of the emitted electrons reaches up to 3 photon energies above the photoemission threshold. Our emission spectra are consistent with models derived for shorter wavelength photoemission [13, 38]. Further studies of this type can lead to new insights about photoemission from nanostructures at intensities between the perturbative and strong-field regimes. An important future measurement is then to investigate the dependence of the photoelectron spectrum on the incident intensity. It is also desirable to improve the stability of the setup in order to better resolve spectral features in the energy-dependent measurements. Both of these future measurements would benefit greatly from an increased repetition rate of the laser system, which would increase the effective dynamic range of the measurement by making imaging at lower peak intensities possible. Finally, because of the technological importance of laser light at 1550 nm for telecommunication purposes, we also believe that this method can be useful in characterizing the near-field properties of optoelectronic components such as converters between optical and electrical signals.

## ACKNOWLEDGMENTS

This work was supported by the Swedish Research Council (VR), the Swedish Foundation for Strategic Research (SSF), the Crafoord Foundation, the Knut and Alice Wallenberg Foundation, the Marie Curie Intra-European fellowship ATTOCO, and the European Research Council (ERC) startup grant ElectronOpera.